# Anomalous selective reflection in cholesteryl oleyl carbonate - nematic 5CB mixtures and effects of their doping by single-walled carbon nanotubes

N.I. Lebovka,[a*] L.N. Lisetski,[b] M.I. Nesterenko,[b] V.D. Panikarskaya,[b] N.A. Kasian,[b] S.S. Minenko,[b] and M.S. Soskin[c]

[a]*Institute of Biocolloidal Chemistry, NAS of Ukraine, 42 Vernadsky Avenue, 03142 Kyiv, Ukraine*

[b]*Institute of Scintillation Materials of STC "Institute for Single Crystals", NAS of Ukraine, 60 Lenin Ave., Kharkov 61001, Ukraine*

[c]*Institute of Physics, NAS of Ukraine, 46 Nauki Prosp, 03022 Kyiv, Ukraine*

Liquid crystalline (LC) mixtures of cholesteryl oleyl carbonate (COC) and 4-pentyl-4'-cyanobiphenyl (5CB), as well as dispersions of single-walled carbon nanotubes (NTs) in these mixtures, were studied by means of selective reflection measurements, differential scanning calorimetry (DSC) and optical microscopy. The relative mass of COC in a mixture $X$ was varied between 0.4 and 1.0, the temperature range of measurements was between 284 K and 314 K, and concentration of NTs was fixed at 0.1 %. Two important anomalies were noted: (1) the cholesteric to smectic-A transition temperature increased on dilution of COC by non-smectogenic 5CB in the concentration range $0.8<X<1$, and (2) the reciprocal pitch vs. 5CB concentration dependence was essentially linear, in contrast to behaviour commonly observed in nematic-cholesteric mixtures. A model of molecular arrangement in the mixtures, accounting for the possibility of integration of 5CB dimers and monomers between COC molecules and presumably explaining the experimental data, was proposed. The helical pitch of the cholesteric mixtures remained practically unchanged upon doping by NTs, and only slight widening of the selective reflection peaks was noted. The obtained results allow considering the COC+5CB mixtures as promising matrices for composite materials on the basis of liquid crystals and carbon nanotubes.

[*] Corresponding author. Tel.: 380 44 4240378; Fax: 380 44 4248078

*E-mail address:* lebovka@gmail.com (N. Lebovka).

## 1. Introduction

Nowadays the suspensions of various micro- and nano-particles in liquid crystals (LC), also known as liquid crystal colloids, steadily attract a great fundamental and practical attention [1–3]. Of special interest are LC suspensions of carbon nanotubes (NTs), which can display many intriguing properties, caused by integration of highly anisotropic particles into orientationally ordered structure of LC [4–6]. While NT suspensions in nematic LC have already demonstrated the possibility of useful practical applications, and in-depth studies of organization of these systems on the molecular level (formation of NT aggregates, percolation phenomena, etc.) have been carried out [5,7–13], only few papers on NT suspensions in cholesteric LC can be found [14–19]. One can mention the preliminary study that demonstrated the effect of cholesteric mixture content on the aggregation and sedimentation stability of NTs [14] or the dielectric studies of NT suspensions in a nematic with a chiral dopant [15]. Effects of CNTs on mesomorphic transition temperatures and viscosity of cholesterol esters were reported in [16], and in [17] the chiral nematics with dispersed NTs were used as a hybrid sensor material for vapor detection by combined monitoring of changes in selective reflection and conductivity. In our previous studies, we observed that optical transmission vs. temperature behavior of CLC+NTs suspensions was rather similar to that of their nematic counterparts, and the effects of NTs on selective reflection spectra were rather weak, so, it was argued that NTs were probably aligned in the planes of quasi-nematic layers [19]. Note, that formation of CNT aggregates could be much slower in the cholesteric phase as compared with similar suspensions in nematics [18].

In this work, we used mixtures of cholesteryl oleyl carbonate (COC) (widely used in cholesteric materials for thermography and other applications [20,21]) and a typical nematic 4-pentyl-4'-cyanobiphenyl (5CB) in a broad range of their concentrations. The systematic data on changes in selective reflection spectra of the said mixtures with variation of temperature and concentrations of the LC system components in the absence and in the presence of dispersed single-walled NTs should give us insight into specific features of intermolecular interaction and supramolecular ordering in such a complex hybrid system.

## 2. Materials and methods

Cholesteryl oleyl carbonate (COC) was obtained from Aldrich, USA. Under cooling, pure COC exhibits the isotropic(I)→cholesteric(Ch) transition at $T_{ICh} \approx 309$ K, cholesteric(Ch) → smectic A (SmA) transition at $T_{ChSm} \approx 295$ K and smectic A (SmA) → solid(C) transition at $T_{SmC} < 273$ K. On heating, the crystallized COC melts above $\approx 295$ K directly into the cholesteric phase, so, the smectic-A phase of COC is classified as monotropic.

The nematic 5CB of 99.5 % purity (Chemical Reagents Plant, Ukraine) undergoes the nematic-isotropic transition at $T_{N \to I} \sim 308 - 309$ K and the crystal-nematic transition (melting) at $T_{C \to N} = 295.5$ K. The chemical structure of COC and 5CB and their phase transitions on cooling are shown in Fig.1.

The single-walled carbon nanotubes (NTs) having $d \sim 1.5$ nm in diameter and $l \sim 5$-10 microns in length were obtained from Arry, Germany. The LC+NT suspensions were prepared by adding 0.1 % (mass) of NTs to the LC solvent in the isotropic state with subsequent 20–30 min sonication of the mixture using a UZDN-20/40 ultrasonic disperser (Ukrrospribor, Sumy, Ukraine), at the frequency of 22 kHz and the output power of 150 W in accordance with procedure essentially similar to the previously described [8,12,19]. After sonication, the suspensions were rapidly cooled to the temperature ~320K.

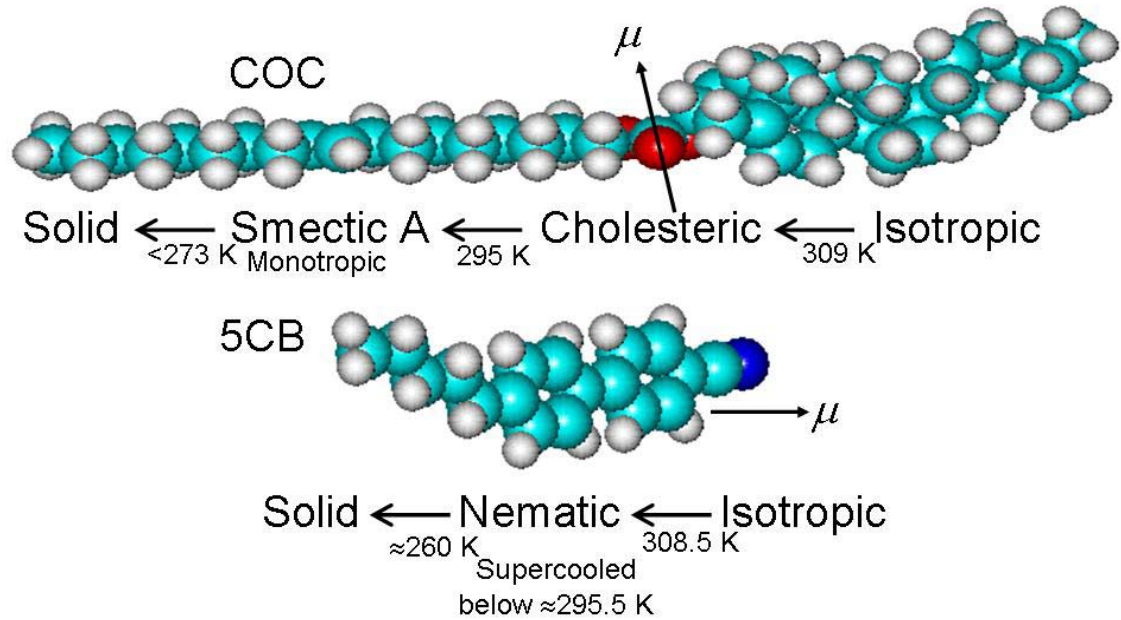

Figure 1. Chemical structure of COC and 5CB molecules. Arrows show the direction of dipole moments.

The relative concentration of COC in COC+5CB mixture, $X$, was determined as
$$X = m_{COC}/(m_{COC}+m_{5CB}) \tag{1}$$
where $m_{COC}$ and $m_{5CB}$ are the masses of COC and 5CB in the mixture.

Selective reflection spectra were measured using a Shimazu UV-2450 (Japan) spectrophotometer within 300-900 nm spectral range. A sandwich-type LC cells (with 20 μm thickness) were used. The cell walls were treated with polyvinyl alcohol water solution and, after drying, rubbed in one direction to obtain the planar texture [19]. The sample was introduced between the cell walls using the capillary forces at the temperature of ~313 K, which corresponds to the isotropic phase. The measurements were done within the temperature range of 290-310 K in the heating and cooling modes, and the temperature was stabilized using a flowing-water thermostat (±0.1 K). For the same cells the optical microscopy images were obtained using OI-3 UHL 4.2 microscope (LOMO, Russia).

The phase transition temperatures were determined from the differential scanning calorimetry (DSC) measurements using a thermoanalytical system Mettler TA 3000 (Switzerland). The measurements were carried out both on heating and cooling; the mass of each sample was ~20 mg, and the scanning rate was 2 K/min, which ensured clear peaks on the thermograms. The heat flux and temperature calibration was done using the metal indium standard at the same scanning rate as in experiments with the tested samples.

## 3. Results and discussion

### *3.1. COC+5CB mixtures*

Figure 2 shows examples of the selective reflection spectra of pure COC (a) and 70% COC+30% 5CB (b) samples with planar texture at different temperatures. During the measurement, the temperature was gradually decreased within the interval corresponding to the cholesteric phase. This resulted in the "red" shift of the selective reflection maximum $\lambda_m$ and increase of the spectral band half-width $\Delta\lambda$.

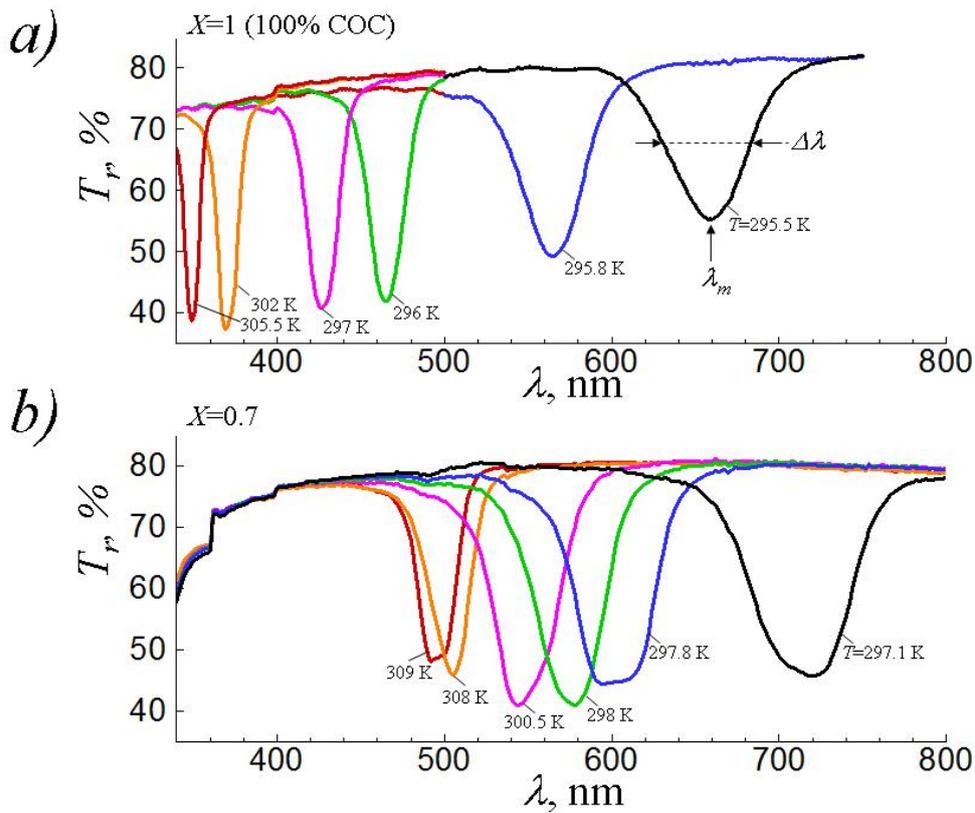

Figure 2. Examples of the transmission, $T_r$, versus wavelength, $\lambda$, dependences for: a) pure COC ($X$=1) and b) COC+5CB mixture ($X$=0.7) at different temperatures. Here, $\lambda_m$ and $\Delta\lambda$ correspond to the selective reflection band maximum and its half-width, respectively.

The obtained temperature dependences of $\lambda_m$ and $\Delta\lambda$ at different values of $X$ (=0.4-1.0) are shown in Fig.3. Behaviour of the pure COC ($X$=1) was typical for cholesteric substances and demonstrated the presence of critical unwinding of the cholesteric helix near the transition temperature between cholesteric and smectic A phases, $T_{Ch-Sm}\approx 295$ K. The divergence of the cholesteric pitch and smearing of the selective reflection bands near the smectic A transition reflected pretransitional formation of local clusters of smectic ordering.

A similar behaviour was observed for COC +5CB mixtures at large content of COC ($X$=0.6-1.0). At smaller concentrations of COC, $X \leq 0.5$, the values of $\lambda_m$ were practically independent of $T$ and the values of $\Delta\lambda$ increased with decreasing temperature in the interval of $T$=296-309 K (Fig.3).

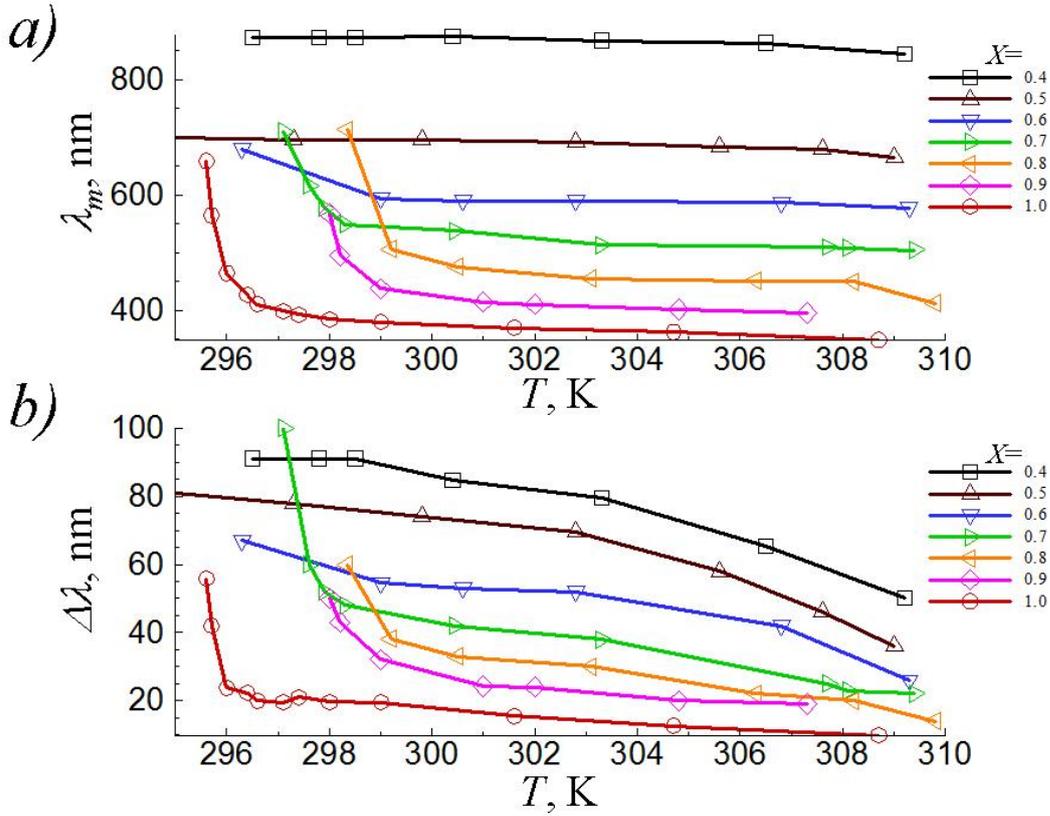

Figure 3. Wavelength of maximum selective reflection $\lambda_m$ (a) and half-width of selective reflection band $\Delta\lambda$ (b) as a function of temperature $T$ for mixtures of COC and 5CB with different concentrations of the components.

*3.1.1. Critical unwinding*

It is interesting to note that in the temperature range close to the region of helix unwinding ($T$≈296-300 K) the values of $\lambda_m$ and $\Delta\lambda$ varied non-monotonously with concentration and displayed the presence of an anomaly. It was reasonable to expect that addition of non-smectogenic 5CB to the COC will decrease the transition temperature $T_{Ch\to Sm}$, and critical unwinding will disappear in the temperature range of measurements on reaching certain 5CB concentration. However, the expected effect of 5CB addition was observed only at $X$<0.8, i.e. at relatively large concentration of 5CB. At $X\geq 0.8$, the initial increase of 5CB content resulted in an increase in the temperature of unwinding $T_{Ch\to Sm}$ (Fig. 3). These results were in full agreement with the obtained DSC data.

Figure 4 presents concentration dependence of cholesteric-smectic A transition temperature $T_{Ch\to Sm}$ and examples of DSC thermograms (insert). The value of $T_{Ch\to Sm}$ passed through a maximum at $X\approx 0.8$. The normal behaviour, i.e. the increase of $T_{Ch\to Sm}$ with $X$, was observed only at $X$<0.8, while the $T_{Ch\to Sm}(X)$ dependence was anomalous at small concentrations of 5CB in COC.

The initial increase of $T_{Ch \to Sm}$ on COC dilution by 5CB within $X$=0.8-1.0 (Fig. 4) may be explained in the same way as the formation of so-called "induced smectics", generally ascribed to a kind of specific interactions between the molecules in the mixture [22–24].

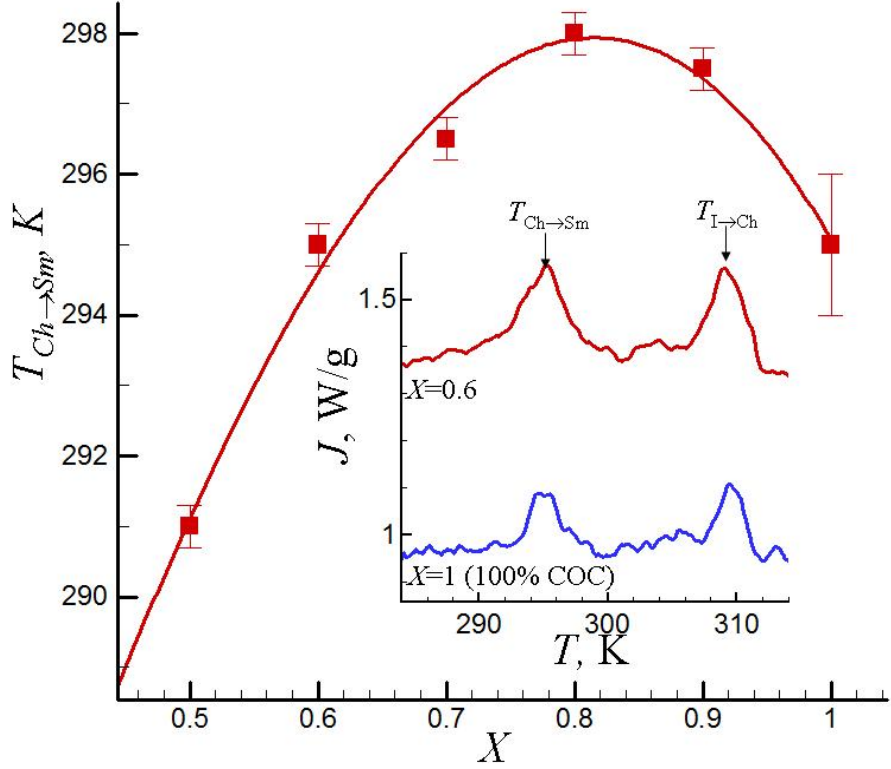

Figure 4. Temperature of the cholesteric-smectic A phase transition of COC-5CB mixtures as a function of 5CB concentration. Insert shows DSC thermograms of pure COC ($X$=1) and COC + 5CB mixture ($X$=0.6) with approximately the same value of $T_{Ch \to Sm} \approx 295$ K.

Note that the temperature of the cholesteric-smectic A transition, $T_{Ch \to Sm} \approx 295$ K, appears to be approximately the same for pure COC ($X$=1) and for COC + 5CB mixture with $X \approx 0.65$. This can explain the observation made in [25,26] that, surprisingly, addition of about 30% 5CB did not change the cholesteric – smectic-A transition temperature.

*3.1.2. Temperature range far from critical unwinding*

The concentration variations of $\lambda_m$ and $\Delta\lambda$ at temperatures far from critical unwinding (i.e., at $T \geq 300$ K) were monotonous (Fig. 5). The observed concentration behaviour of $\Delta\lambda$ at different fixed temperatures (Fig. 5b) was quite natural and could be explained by the fact that birefringence $\Delta n$ of 5CB is much larger as compared to that of COC.

The analysis has shown that the $1/\lambda_m$ versus $X$ dependence was practically linear

$$1/\lambda_m = X/\lambda_o \qquad (2)$$

where, $\lambda_o \approx 355\pm3$ nm is $\lambda_m$ value for pure COC($X=1$), and determination coefficient is 0.998 (Fig. 5a).

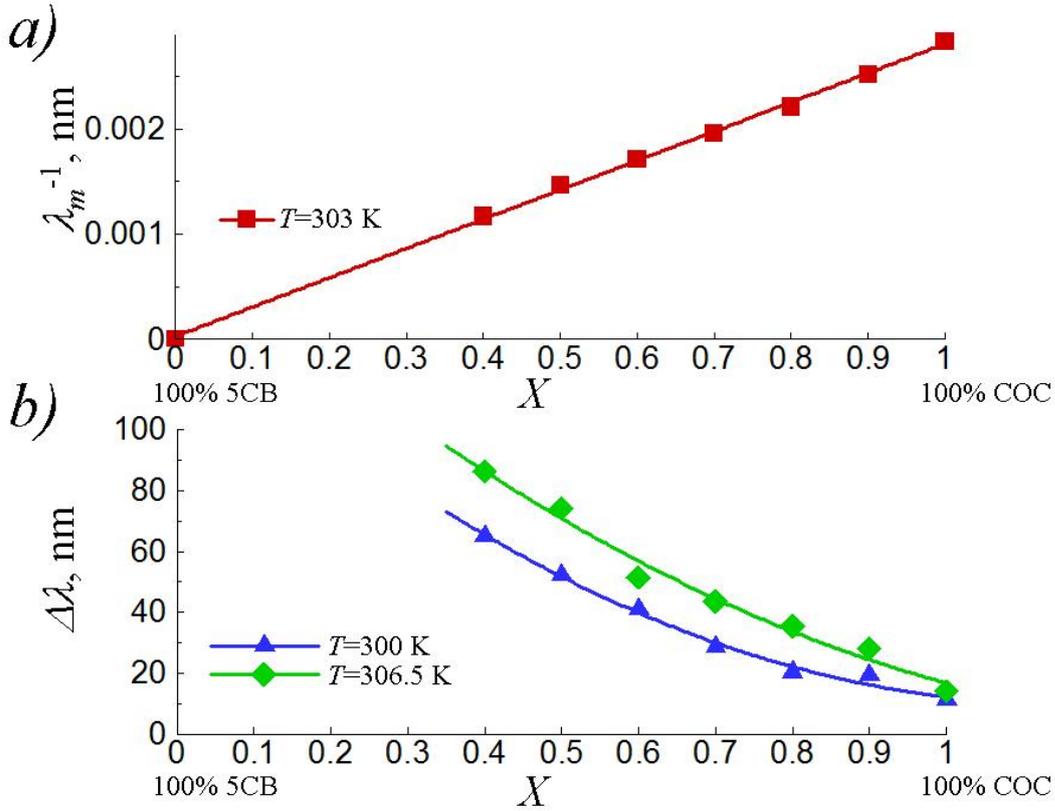

Figure 5. The value of $1/\lambda_m$ (a) and half-width of the selective reflection peak $\Delta\lambda$ (b) as a function of 5CB concentration in COC-CC mixtures at different fixed temperatures.

It has been known that linear $1/\lambda_m$ versus $X$ dependences are typical for mixtures with chemically similar (or almost similar) nematic and cholesteric components [27]. As for mixtures of nematics with cholesterol esters, this dependence is typically non-linear, with deviations from linearity corresponding to extra twisting [28]. This anomaly can be explained by certain specific interactions between 5CB and COC molecules. The presence of such interactions in COC+5CB mixtures is supported also by the above-described anomalous concentration behaviour of $T_{\text{Ch}\to\text{Sm}}$ at small dilution of COC by 5CB (at $X\geq0.8$).

*3.1.3. Model of mutual arrangement of COC and 5CB molecules*

It can be speculated that the following model of mutual arrangement of COC and 5CB molecules is true (Fig.6). Several experimental studies have indicated dimerization of 5CB in a nematic phase [29–31]. The calculations show that the most favourable arrangement of 5CB molecules in a dimer–like pairs (dimers) corresponds to the antiparallel side-by-side geometry [32,33]. Note that the mean diameter of a single 5CB molecule in the direction perpendicular to its long axis is 0.43 nm and its length is 1.87 nm [34]. However, 5CB molecules associate in a head-to-tail manner forming the antiparallel dimers with thickness estimated as 2.5 nm [35,36]. At small concentrations of COC, the dimers can be easily located between long hydrocarbon chains of adjacent

COC molecules, thus effectively enhancing translational ordering and stabilizing the smectic-A phase (Fig. 6a).

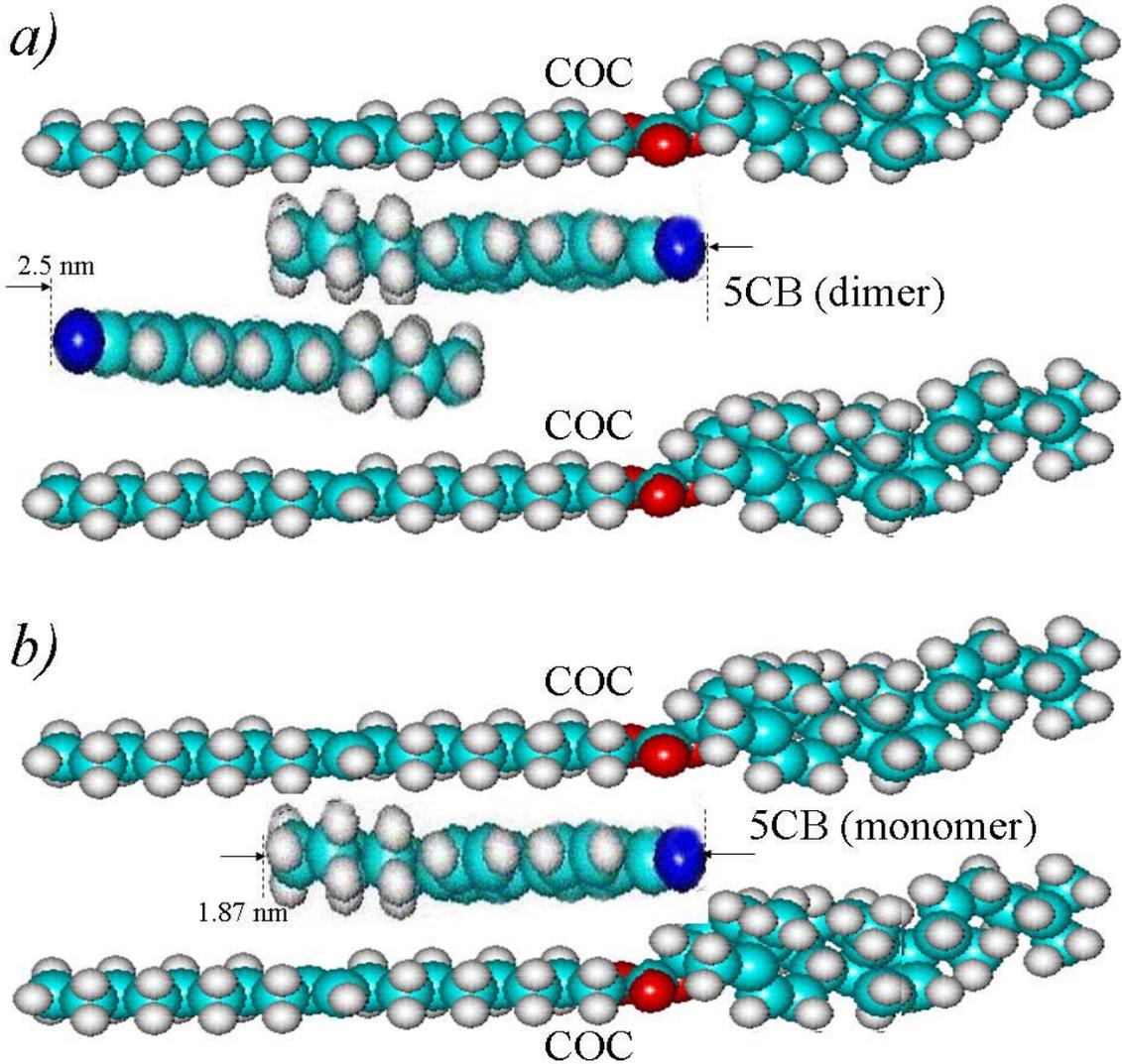

Figure 6. Possible models of integration of 5CB dimer-type pairs (a) and monomeric 5CB molecules (b) between COC molecules.

However, when content of COC ($X \geq 0.8$) is large, the complexing interactions between 5CB and COC molecules and destruction of 5CB dimers are possible. The similar effects were reported earlier for mixture of 5CB with smectic EPAB (4-ethyl-4'-n-pentylazoxybenzene) [29]. The observed anomalous extra twisting of cholesteric helix and increase of $T_{Ch \to Sm}$ at initial dilution of COC by 5CB may reflect complexing interactions of 5CB and COC molecules (Fig.6b). The critical concentration of $X \approx 0.8$ corresponds to destruction of 5CB dimers in the concentrated solution of COC.

### *3.2. Suspensions of NTs in COC+5CB mixtures*

Finally, the effects caused by introduction of single-walled carbon NTs into the COC+5CB mixtures were studied. Introduction of NTs (0.1%) into the pure COC resulted in a decrease in the transition temperature $T_{Ch \to Sm}$ and the relatively small changes of $\lambda_m$ (Fig. 7a) and $\Delta\lambda$ (Fig. 7b) near the temperature of critical unwinding. This can indicate that under conditions of our experiments NTs were strongly integrated into the structure of pure COC.

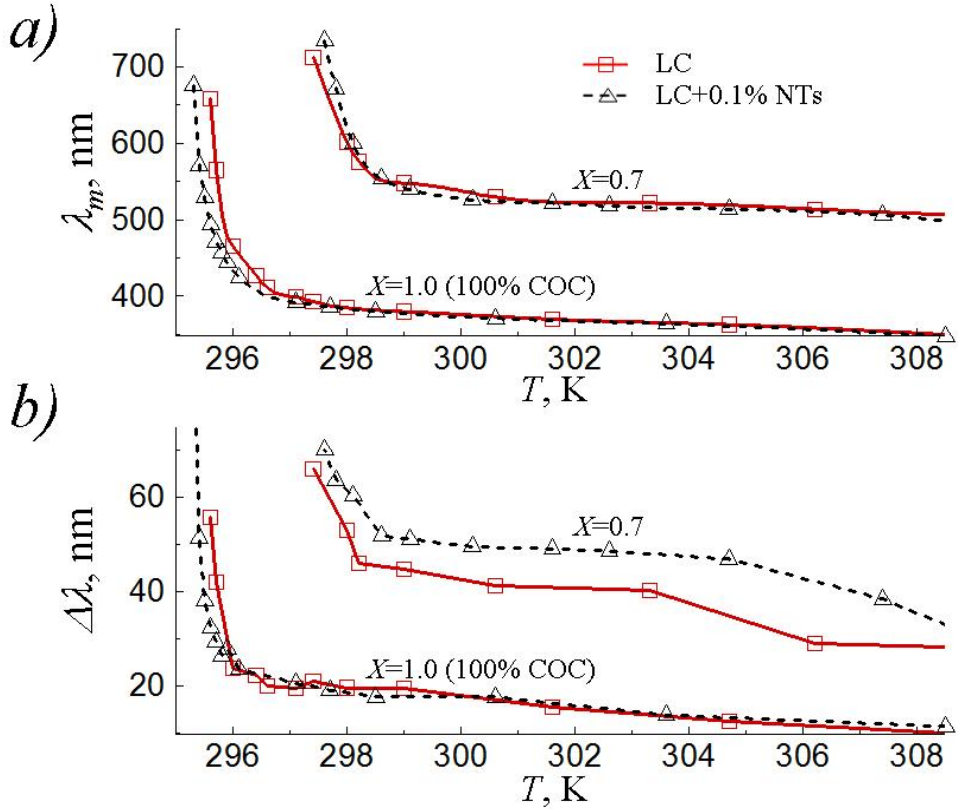

Figure 7. Wavelength of maximum selective reflection $\lambda_m$(a) and half-width of the selective reflection peak, $\Delta\lambda$ (b) as a function of temperature for $X$=0.7 and $X$=1.0 without and with dispersed single-walled carbon nanotubes.

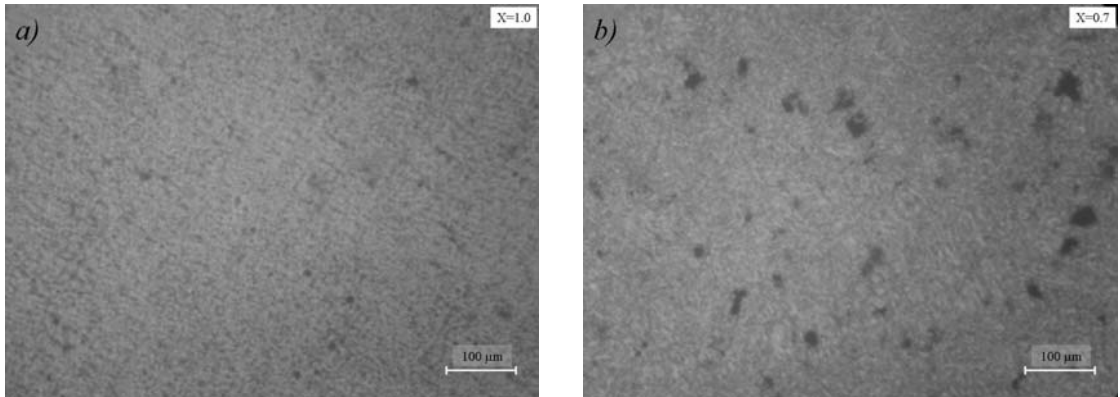

Figure 8. Micro-photos of 0.1% NT suspensions in COC (*a*, $X$=1.0) and COC+5CB mixture (*b*, $X$=0.7). $T$≈300 K.

From the other side, introduction of single-walled NTs (0.1%) into COC+5CB mixture ($X$=0.7) resulted in the opposite effects. The NTs facilitated unwinding of the spirals (Fig. 7a) and caused noticeable broadening of the selective reflection band (Fig. 7b) near the region of critical unwinding. These effects can reflect changes in the degree of NT integration into the COC+5CB mixture.

Analysis of the microstructure of NT suspensions justified different integration of NTs in COC+5CB mixtures depending on $X$. Figure 8 presents micro-photos of 0.1% NT suspensions in COC (*a*, $X$=1.0) and COC+5CB mixture (*a*, $X$=0.7). The detailed experiments have shown that the best possible dispersing of NTs was always observed in pure COC; i.e., COC is a "good" solvent for NTs. The maximum size of NT aggregates in COC didn't exceed ≈5-10 µm (Fig.8a). However, the size of aggregates noticeably increased with dilution of COC by 5CB (i.e., with decrease of $X$): e.g. it reached 50 µm at $X$=0.7(Fig.8b). Note that such large aggregates of NTs may cause broadening of the selective reflection band in the COC+5CB mixture (Fig. 7b). It means that 5CB is a "bad" solvent for NTs and the NTs exhibit strong tendency to formation of clusters with introduction of 5CB. It is in agreement with the earlier observed aggregation of multi-walled carbon nanotubes in 5CB [12].

Note that multi-walled NTs with large outer diameter (8−15 nm) can produce macroscopic helical twisting even in achiral nematic 5CB [37]. It was explained by manifestation of inherently chiral properties of NT surface and it appears that this effect was essential in our case because of the small diameter of single-walled NTs (d ~ 1.5 nm) used in this work.

## 4. Conclusions

Different anomalous properties were demonstrated by the liquid crystalline mixtures of long-chained cholesterol ester (COC) and nematic cyanobiphenyl (5CB).

(1) (1) "Stabilisation" of the monotropic SmA phase by 5CB, similar to formation of an "induced smectic", was observed close to the temperature of critical unwinding of the cholesteric spiral, $T_{Ch \to Sm}$, at high concentrations of COC (0.8≤$X$≤1.0). The maximum of $T_{Ch \to Sm}$ ≈298 K was reached at extremum concentration, $X$≈0.8, and further decrease of $X$ below 0.8 resulted in decrease of $T_{Ch \to Sm}$.
(2) At temperatures far from the temperature of critical unwinding, the observed dependence of reciprocal pitch 1/$\lambda_m$ vs. COC concentration $X$ was linear, which was also surprising. Such type of dependence can be expected for mixtures with chemically similar nematic and cholesteric components (they can differ from each other only by the presence of an optically active centre) [27]. It was not the case for COC and 5CB molecules, and explanation of this anomaly requires supplementary clarifications.
(3) The proposed model accounts for the possibility of integration of 5CB dimer-type pairs (at $X$<0.8) and monomeric 5CB molecules (at $X$>0.8) between COC molecules.
(4) COC+5CB mixtures can be considered as promising matrices for composite materials on the basis of LC and NTs. Doping of such mixture by NTs (0.1%) resulted in the relatively small changes of $\lambda_m$ and $\Delta\lambda$. The observed changes reflected the state of NT dispersing that was noticeably improved when concentration of COC, $X$, was increased.